\documentclass[%
 reprint,
showpacs,
 amsmath,amssymb,
 aps,
 pra,
 longbibliography,
 floatfix,
 superscriptaddress
]{revtex4-2}

\usepackage[utf8]{inputenc}	
\usepackage[T1]{fontenc}	
\usepackage[]{graphicx}		
\usepackage{xcolor}
\usepackage{gensymb}
\usepackage{amsmath, mathdots}
\usepackage{braket}
\usepackage{comment}
\usepackage[english]{babel}
\usepackage{color}
\usepackage{soul}

\renewcommand{\H}{\hat{H}}
\renewcommand{\L}{\hat{L}}
\renewcommand{\a}{\hat{a}}
\renewcommand{\b}{\hat{b}}
\renewcommand{\c}{\hat{c}}
\renewcommand{\d}{\hat{d}}

\newcommand{\gin}[1]{g_{\textup{in},#1}}
\newcommand{\gout}[1]{g_{\textup{out},#1}}

\begin{document}

\title{Interactions in Quantum Networks with Pulse Propagation Delays}

\author{Victor Rueskov Christiansen}
\email{victorrc@phys.au.dk}
\affiliation{Center for Complex Quantum Systems, Department of Physics and Astronomy, Aarhus University, Ny Munkegade 120, DK-8000 Aarhus C, Denmark}

\author{Klaus M{\o}lmer}
\email{klaus.molmer@nbi.ku.dk}
\affiliation{Niels Bohr Institute, University of Copenhagen, Blegdamsvej 17, 2100 Copenhagen, Denmark}

\date{\today}

\begin{abstract}
\noindent 
The finite speed of light of pulses implies propagation delays between the generation of light, the interaction with different elements along its path and the final detection at the output of networks and interferometers. We show that it is possible to take these delays into account with a theoretical method which evades quantization of the full continuum of field modes. We illustrate the use of this method by analyzing Ramsey excitation of an atom by a split and delayed quantum pulse. 
\end{abstract}

\maketitle
\noindent

\section{Introduction}

Quantum networks are spatial architectures where propagating quantum fields enable the transfer of states, excitation, or entanglement between different nodes. Such networks may display novel entanglement properties \cite{gisin-entanglement}, and they may serve practical purposes as clocks \cite{sorensen-ye-clocks}, sensors \cite{pezze-sensor, zhuang-sensing} and form the basis of secure quantum communication \cite{epping-qkd} and distributed optical quantum computing \cite{grover1997quantumtelecomputation, cirac-dqc, main-dqc}.

The fact that propagating quantum fields occupy a continuum of modes, e.g., wavenumber eigenmodes or coupled space-bin modes, poses a challenge to their theoretical description. This challenge can be alleviated when propagation times are much shorter than the natural time scale of physical interactions, in which the interaction with multiple components can be treated as simultaneous. If an arbitrarily extended network permits only uni-directional and non-dispersive propagation, one can absorb propagation delays in the synchronization of clocks at each node, such that also in this case, the interactions appear formally simultaneous. It is subsequently possible to employ the Born-Markov approximation to eliminate the field modes altogether and obtain coupled master equations of the systems located at the network nodes. 

Such equations were introduced by Gardiner \cite{gardiner-input-output, gardiner-cascade} and Carmichael \cite{carmichael-cascade} and were referred to as cascaded master equations to reflect the fact that the output from one node serves as input to other ones in the network. A general, so-called, SLH formalism, named after matrix components, S, L and H, provides an efficient determination of the master equation for general architectures of concatenated network elements \cite{SLH-Gough, SLH-framework}.

\begin{figure}
    \centering
    \includegraphics[width=0.9\linewidth]{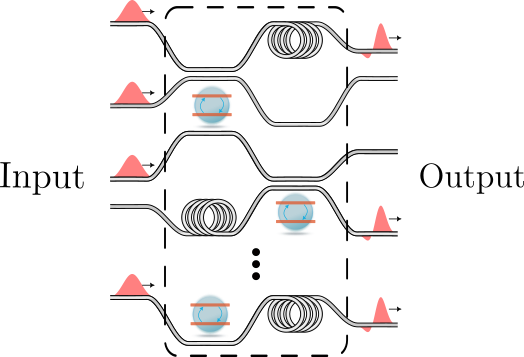}
    \caption{An example of a quantum network with several input and output quantum channels with integrated delay lines and non-linear components illuminated by pulses of quantum radiation.}
    \label{fig:example-network}
\end{figure}

When the assumption of negligible time delay between the interaction with different components breaks down or cannot be remedied by a local synchronization of time, such as in the example network in figure \ref{fig:example-network} where delay lines are an integral part of the network dynamics, the efficient cascaded master equation does not apply, and more complex methods are needed. We note in particular the quantization of the propagating field across discrete bins in space or time, which explores a Hilbert space of vast dimension but does permit an effective treatment by matrix product states. This approach has been implemented to account for propagation and data processing delays in the interesting case where measurements of output signals are used to guide feedback operations on the emitters (delayed with respect to the time of emission) \cite{Pichler-MPS-delay, Grimsmo-time-delay, Whalen-Grimsmo-time-delay, waveguide-qed-jl}. Another approach, which is more similar to the approach we take here, is to introduce fictitious elements that model the effect of time delays to some perturbative order \cite{Stace-dispersion-cavity, Tabak-passive-delay}.

In many applications of quantum networks, the focus is on the quantum states held by the nodes and carried by propagating wave packets of radiation. A wave packet is a single mode, and its quantum state is analogous to the states of a single-mode cavity field, for which interactions with atomic systems have been studied in large detail. An important difference between cavity eigenmodes and propagating fields, however, is that the former have discrete frequencies and only one mode may couple resonantly to a given emitter, while propagating fields explore a mode continuum. Wave packets are hence both subject to dispersion and entanglement between the number of quanta and their spatio-temporal profile by non-linear scatterers. The latter may be seen as a loss of (single-mode) fidelity, but also as a resource for genuine multi-mode functionalities \cite{photon-sorting, photon-subtraction}. Formally the joint evolution of the photon number statistics and propagation of the wave packet can be taken into account by an effective  formalism that represents input travelling pulses as the  output from virtual cavities and subsequently determines the multi-mode character of the output fields \cite{short_kiilerich, long_kiilerich, interaction-picture,virtual-cavity-review}. 
      
In this article, we are interested in how propagation delays affect the interaction of pulses of light with scattering elements. With a focus on the physics of single-mode pulses, we propose to represent the time delays by effectively capturing a single or few wave packets in virtual cavities and releasing them with the desired delay. Unlike the discretisation in time bin modes, we hence only need one single auxiliary mode to describe the time delay of a pulse, irrespective of the duration of the delay, while enjoying a non-perturbative approach in contrast to the method of using fictitious elements.

The article is structured as follows. In Section \ref{sec:delaying-theory} we present the virtual cavity method for delaying pulses of quantum radiation in a network. In Section \ref{sec:Mach-Zehnder} we discuss a simple example of how to use the method to delay the field in one channel in a network with respect to another one and we present a method to reduce the complexity that arises from using the method presented in Sec. \ref{sec:delaying-theory}. In Section \ref{sec:Ramsey-spectroscopy} we employ our method to analyse illumination of an atom by a radiation pulse that has been split at a beam splitter and its components have been delayed with respect to each other. This is the central process in Ramsey spectroscopy, where the use of classical, coherent pulses of radiation are treated by a simple atomic master equation. We generalize this treatment to illumination by entangled pulses of quantum radiation. In section \ref{sec:non-linear-output} we analyse the intensity of the output radiation from the atom. In the conclusion and outlook, we indicate applications and natural next directions of research building on the pulse quantization and virtual cavity approach.

\section{A theory model for delayed pulses of radiation} \label{sec:delaying-theory}

In experiments, pulses of light are delayed if they are sent along extended optical paths. This article is not about experimental schemes to generate time delays, but about the theoretical description of interactions between quantum systems and pulses of light in networks with inherent time delays. Rather than employing a quantized version of Maxwell's equations for the propagating field, the delay of the time of arrival of a quantum pulse due to a propagation distance can be simulated by the capture and subsequent release of the pulse. For our purpose, we shall simulate pulse delays by the equivalent process of storing the pulse in a cavity mode and releasing it at a later time. 

In \cite{short_kiilerich, long_kiilerich, virtual-cavity-review, interaction-picture} virtual cavities were similarly used to model the launching of a quantum pulse with pulse shape $u(t)$ onto any downstream system, by assuming the initial quantum state of the pulse is prepared in an eigenmode of a cavity, which is coupled to the environment with the time dependent coupling strength,
\begin{equation} \label{eq:gu}
    \gout{u}(t) = \frac{u^*(t)}{\sqrt{1 - \int_0^t dt' |u(t')|^2}}.
\end{equation}
Similarly, to perfectly transfer radiation occupying a pulse with the shape $u(t)$ into a cavity eigenmode, the coupling to this cavity must be
\begin{equation} \label{eq:gv}
    \gin{u}(t) = \frac{-u^*(t)}{\sqrt{\int_0^t dt' |u(t')|^2}}.
\end{equation}

\begin{figure}
    \centering
    \includegraphics[width=\linewidth]{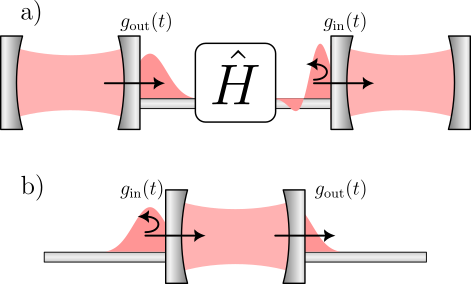}
    \caption{Virtual cavity approach to pulses of quantum radiation. (a) A pulse mode $u(t)$ in any desired quantum state can be theoretically modelled as the output from a single-mode cavity. The interaction with a quantum node and the quantum state contents of any downstream wave packet $v(t)$ can be obtained by solving a cascaded master equation for the emitting $u$-cavity, the quantum node and an appropriate $v$-cavity. (b) A pulse $v(t)$ can be delayed by first storing it in a cavity mode and subsequently releasing it, described formally by the same processes as shown in panel (a). Pulses orthogonal to the chosen one will be reflected by the delay cavity and will not be delayed, unless other designated delay cavities are employed to that effect.}
    \label{fig:delay_cavity}
\end{figure}

We shall employ the same formalism to model the delay of a quantum pulse by its absorption in a virtual cavity and its re-emission at a chosen later time, as pictured in figure \ref{fig:delay_cavity}. 
The pulse is first given by $v(t)$ and is absorbed by the virtual cavity with coupling rate $\gin{v}(t)$. Hereafter, the cavity coupling rate of the output port follows $\gout{u}(t)$ to emit the pulse in the desired shape $u(t)$. If only a delay is desired, we choose $u(t) = v(t-\tau)$, where $\tau$ is the desired time delay. Note that this step-wise approach functions only if $\tau$ is large enough that the pulse can be fully captured before it is re-released. In the next sections, we will discuss modifications of the theory scheme that permit treatment of arbitrarily short delays. 

In the same way, several simultaneously occupied pulses of radiation can all be delayed by storage in and release from a cascaded array of virtual cavities with appropriate time-dependent coupling strengths \cite{short_kiilerich, long_kiilerich}. It may seem that employing an auxiliary oscillator mode for the delay of each pulse will extensively complicate the calculation of the dynamics. But, we will show in the following that since the delaying dynamics is linear, an effective interaction picture applies and simplifies the  calculations considerably  \cite{interaction-picture, virtual-cavity-review}.

\subsection{Delaying a single-mode field by an arbitrary amount}
Fully absorbing and then reemitting pulses from a virtual cavity is incompatible with delays $\tau$ shorter than the pulse duration. For such short delays, we assume a single, time-dependent cavity coupling that ensures the desired balance between the input and output amplitudes at any time while storing part of the field inside the cavity for later release. We note that in eqs. \eqref{eq:gu} and \eqref{eq:gv}, the denominator represents the fraction of the pulse residing inside the emitting and absorbing cavities, respectively. To account for the fact that the cavity is fed by an incoming pulse, $v(t)$, while at the same time radiating an outgoing pulse, $u(t)$, we simply change $1$ in the denominator of eq. \eqref{eq:gu} to the integrated radiation that has entered the cavity so far. This means we get the out-coupling strength 
\begin{equation} \label{eq:gu_modified}
    \tilde{g}_\textup{out,u,v}(t) = \frac{u^*(t)}{\sqrt{\int_0^t dt' |v(t')|^2 - \int_0^t dt' |u(t')|^2}},
\end{equation}
where $v(t)$ is the shape of the arriving pulse being gradually absorbed while the cavity also leaks a pulse in the shape of $u(t)$. At the same time, modifying the assumption that the cavity does not lose any radiation while it picks up mode $v(t)$ implies that the incoupling strength, through the input mirror takes the value, cf., eq. \eqref{eq:gv},
\begin{equation} \label{eq:gv_modified}
    \tilde{g}_\textup{in,v,u}(t) = \frac{-v^*(t)}{\sqrt{\int_0^t dt' |v(t')|^2 - \int_0^t dt' |u(t')|^2}}.
\end{equation}
With these expressions, we can convert radiation between pulse shapes as long as the integrated emitted power is at any time less than or equal to the integrated incident power. We are interested in the simple delay $u(t) = v(t - \tau)$, and we observe that when $\tau$ is larger than the duration of $v(t)$, eqs. \eqref{eq:gu_modified} and \eqref{eq:gv_modified} take the form of eqs. \eqref{eq:gu} and \eqref{eq:gv} in two consecutive time intervals.

\subsection{Delaying a multi-mode field}
We have not found a way to use the same method and a single cavity to delay fields containing multiple pulse modes, and instead, we have recourse to sequentially fully absorb all relevant input-pulses $v_i(t)$ in separate cavities, before reemitting them a time $\tau + T$ later where $\tau$ is an adjustable relative delay and $T$ is the time it takes to fully absorb the pulses.

In this case, it is hence also not possible to delay the pulses by an arbitrarily small amount, as the smallest delay possible is $T$. However, in most situations, it is not the actual delay that is important, but rather the relative delay between different propagation channels, as in an interferometer. In such cases, we may capture and store the pulses in all quantum channels, and release them again, with any desired relative delays. In the next section, we will show this for a Mach-Zehnder interferometer with different path lengths. For simplicity, we consider delaying only a single pulse, but the method may be extended to a multimode field by including several delay cavities.

\section{Mach-Zehnder interferometer with different path lengths} \label{sec:Mach-Zehnder}
In Mach-Zehnder interferometers, one optical path is longer than the other, and an incident pulse may be split and the components may be delayed with respect to each other. In the continuous wave regime, this would be described by a phase-shift of the radiation in one path, while for pulses of finite time length, the effect is more complicated. 

To simulate the relative delay between two paths, we use the method described in the previous section, assuming virtual delay cavities along both paths to pick up the radiation that is scattered on the beam-splitter. Once the radiation has been picked up in both paths at time $T$, the radiation is reemitted first from the delay cavity representing the shorter path and, after a relative delay $\tau$, by the cavity representing the longer path. With this method, we can model any path length difference. Note that while pulse scattering on a linear beam-splitter may be solved by other means, our method is more general and also applies to scattering off non-linear systems. We give an example of this in section \ref{sec:non-linear-output}.

It seems to be a significant disadvantage of our method that it generally needs two oscillators to delay one pulse relative to another one, and the Hilbert space dimensionality has hence increased drastically to account for the delay. Since, however, the input field is single mode and the beam-splitter merely performs a mode transformation, the system remains single mode in the proper basis. Furthermore, we can simplify the linear dynamics by going to an appropriate interaction picture.

\begin{figure}
\includegraphics[width=\linewidth]{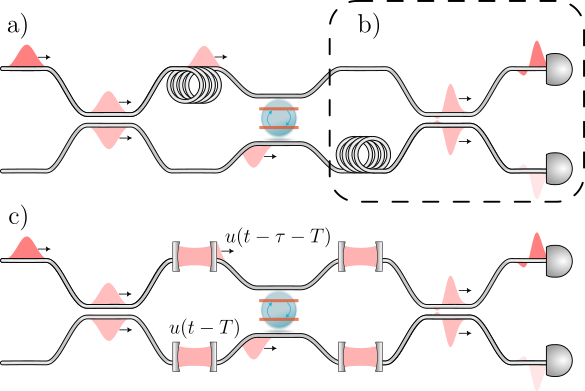}
\caption{Ramsey-like interferometry experiment. In panel (a) a pulse of quantum radiation is split on a beam-splitter before entering a long and a short path. Both paths pass in the vicinity of a two-level system, which interacts with the pulses of quantum radiation. In panel b) the outputs after interaction with the two-level system are further delayed to be superimposed after the same total delay on a beam-splitter and subsequently detected by photo-detectors placed in both paths. In panel c) the difference in path length is modelled by virtual delay cavities that emit the pulses with a relative delay $\tau$ to emulate the effect of the different path lengths in the physical network.}
\label{fig:Ramsey}
\end{figure}

We consider a setup where a virtual cavity is used to describe the pulse inbound on a beam splitter with pulse shape $u(t)$. To simulate the delay between the paths of the interferometer, a delay cavity is placed in each path designed to pick up the radiation in the same mode $u(t)$ and reemit it at a later time in modes $v_1(t) = u(t - \tau - T)$ and $v_2(t) = u(t - T)$, where $T$ is chosen such that both cavities have picked up the entire incoming $u(t)$-mode, and $\tau$ is the relative delay of the long path compared to the short path. The Hamiltonian for this interaction is
\begin{align}
\begin{split}
    \hat{H} = \frac{i}{2\sqrt{2}}&\left(\gout{u}(t)\gin{u}^*(t) \a_u^\dagger \a_{v_1} \right. \\ &\left.+ \gout{u}(t) \gin{u}^*(t) \a_u^\dagger \a_{v_2} - \text{H.c.}\right)
\end{split}
\end{align}
where $\a_u$ is the annihilation operator for the virtual input cavity, $\a_{v_1}$ and $\a_{v_2}$ are the annihilation operators for the delay cavities representing the long and short path respectively, and H.c. denotes the Hermitian conjugate. The Hamiltonian evolution is accompanied by two dissipation operators for the two output channels, that reflect the input field
\begin{align}
    \L_{r1} &= \frac{1}{\sqrt{2}} \gout{u}^*(t) \a_u + \gin{u}^*(t) \a_{v_1}, \\
    \L_{r2} &= \frac{1}{\sqrt{2}} \gout{u}^*(t) \a_u + \gin{u}^*(t) \a_{v_2}.
\end{align}
and further two dissipation operators that carry the delayed field
\begin{align}
    \L_{t1} &= \gout{v_1}^*(t) \a_{v_1}, \\
    \L_{t2} &= \gout{v_2}^*(t) \a_{v_2}.
\end{align}
We go to the interaction picture with respect to the Hamiltonian, which describes the transfer of excitation from the input cavity to the delay cavity before the pulse is re-released. We define the operators $\b_1 = (\a_{v_1} + \a_{v_2})/\sqrt{2}$ and $\b_2 = (\a_{v_1} - \a_{v_2})/\sqrt{2}$. In the interaction picture, the time evolution of the operators are
\begin{align}
    \frac{d}{dt}\a_u &= \frac{1}{2} \gout{u}(t) \gin{u}^*(t) \b_1, \\
    \frac{d}{dt}\b_1 &= -\frac{1}{2} \gout{u}^*(t) \gin{u}(t) \a_u, \\
    \frac{d}{dt}\b_2 &= 0.
\end{align}
The solution to these equations is found in \cite{virtual-cavity-review, interaction-picture} for equal input and output pulses and is given by
\begin{align}
    \a_u(t) &= \cos\theta(t) \a_u(0) - \sin\theta(t) \b_1(0), \\
    \b_1(t) &= \cos\theta(t) \b_1(0) + \sin\theta(t) \a_u(0),
\end{align}
where $\theta(t)$ is defined by the relation
\begin{equation}
    \sin^2\theta(t) = \int_0^t dt' |u(t')|^2.
\end{equation}
The Hamiltonian in the interaction picture thus vanishes, $\H_I = 0$, as the interaction picture incorporates the full Hamiltonian dynamics. The dissipation operators simplify,
\begin{align}
    \L_{r1,I} &= \frac{1}{\sqrt{2}} \left(-2 u(t)\csc{(2\theta)} \b_1 + \gin{u}^*(t) \b_2 \right),\\
    \L_{r2,I} &= \frac{1}{\sqrt{2}} \left(-2 u(t)\csc{(2\theta)} \b_1 - \gin{u}^*(t) \b_2\right).
\end{align}
and 
\begin{align}
    \L_{t1,I} &= \frac{1}{\sqrt{2}} \gout{v_1}^*(t) (\cos{\theta}\b_1 + \sin{\theta} \a_u + \b_2), \\
    \L_{t2,I} &= \frac{1}{\sqrt{2}}\gout{v_2}^*(t) (\cos{\theta} \b_1 + \sin{\theta} \a_u - \b_2).
\end{align}
As the incoupling to the delay cavities is designed to fully absorb the incoming pulse shape and reflect only vacuum, we can omit terms with $\L_{r1,I}$ and $\L_{r2,I}$ from the master equation, as they will have no effect, since no radiation leaves along these channels. Furthermore, since the emission of the delayed modes governed by $\gout{v_i}(t)$ happens after the loading of the pulse into the delay cavities governed by $u(t)$, we can replace the trigonometric functions by their long time limits, $\cos\theta \rightarrow 0$ and $\sin\theta \rightarrow 1$ in $\L_{t1,I}$ and $\L_{t2,I}$. We observe then that the $\b_1$-operator disappears in this interaction picture.
The dynamics thus only introduces one additional oscillator mode to account for the relative delay of the incoming pulses. In the next section we will release the two pulses onto a two-level emitter and thereby investigate a Ramsey type interaction of quantum pulses of light with an atom.

\section{Exposing a non-linear scatterer to delayed quantum pulses} \label{sec:Ramsey-spectroscopy}
So far, our analysis and methods concerned only linear propagation and mode mixing, which may be readily solved in the Heisenberg picture.
In this section, we consider a non-linear scattering process with the setup sketched in figure \ref{fig:Ramsey}a), where a two-level system is illuminated by two pulse components with a time delay between them. This constitutes a quantum version of Ramsey spectroscopy, and we aim to show the effect of coherences between the pulses on the atom when the light-pulses have no mean field amplitude but are entangled with each other.

The Hamiltonian for the interaction is
\begin{align}
\begin{split}
    \H = \Delta\hat{\sigma}_+ \hat{\sigma}_- + \frac{i}{2} \sqrt{\frac{\gamma}{2}}&\left(\gout{v_1}(t) \c_+^\dagger \hat{\sigma}_- \right.\\
    &\left.+ \gout{v_2}(t) \c_-^\dagger \hat{\sigma}_- - \text{H.c.}\right),
\end{split}
\end{align}
where $\hat{\sigma}_\pm$ are operators associated with the creation(destruction) of a quantum of excitation in the two-level system, $\Delta$ is the detuning between the incoming radiation and the energy levels of the two-level system, while we choose the two-level atom to have a decay rate $\gamma/2$ along each quantum channel, resulting in a total decay rate $\gamma$. We have furthermore introduced the operators $\c_{\pm} = (\a_u \pm \b_2)/\sqrt{2}$ to ease the notation. The Hamiltonian is accompanied by the two dissipation operators
\begin{align}
    \L_1 &= \gout{v_1}(t) \c_+ + \sqrt{\frac{\gamma}{2}} \hat{\sigma}_-, \\
    \L_2 &= \gout{v_2}(t) \c_- + \sqrt{\frac{\gamma}{2}} \hat{\sigma}_-.
\end{align}

We now turn to integrating the corresponding master equation
\begin{equation}
    \frac{d\rho}{dt} = -i[\H(t),\rho ]+\sum_{i=1}^2 D[\L_i(t)]\rho,
\end{equation}
where $D[\L]\rho \equiv  -\frac{1}{2}(\L^{\dagger}\L\rho + \rho\L^{\dagger}\L) + \L \rho \L^{\dagger}$. We integrate the system numerically using the \texttt{QuTiP} toolbox \cite{qutip1, qutip2}. To investigate if Ramsey-like interference occur when the atom interacts with a Fock state split into an early and late pulse, we vary the detuning $\Delta$ while keeping the relative delay $\tau$ between the pulses constant. We assume an input pulse with a Gaussian temporal profile given by
\begin{equation}
    u(t) = \frac{1}{\sqrt{t_w \sqrt{\pi}}} \exp\left(-\frac{(t - t_p)^2}{2 t_w^2}\right),
\end{equation}
where $t_p$ is the time at which the pulse peaks and $t_w$ is the width of the pulse. For the Ramsey sequence, we choose the relative delay between the pulses to be $\tau = 0.5\gamma^{-1}$ and the width of the pulses $t_w = \pi^{3/2} / (8\gamma n)$, which is the pulse length that gives a $\pi/2$ rotation of the two-level system by a field with mean photon number $n$ in the classical limit. To explore different regimes of photon numbers, we investigate two cases with $n=9$ and $n=30$.

Figure \ref{fig:ramsey-photonic-intensity} shows the atomic population as a function of detuning. Since the atom experiences decay on the same timescale as the pulse width, we plot the excited state population at a time equal to two pulse widths after the peak of the second pulse $t_1 = t_p + \tau + 2t_w$, where the atom has not fully decayed yet. We notice a similar behaviour as in the classical Ramsey effect, where the detuning dependence of the excited state population in the final state comes from the interference between two excitation pathways, $g \rightarrow g \pm e  \rightarrow e$. 
In our calculation, a Fock state $\ket{n}$, which exhibits no classical coherence, drives the atom. Here the binomial splitting of the Fock state on the beam-splitter leads to a coherent superposition of Fock product state components. It is then the interference between pathways that have exchanged and not exchanged a photon between the early and late pulse via the atom that results in the Ramsey-like interference pattern. 

\section{Investigating the output of the non-linear scatterer} \label{sec:non-linear-output}
Having performed the interaction with a non-linear two-level system, we may be interested in the output quantum state of the scattered light, for instance through the interferometer depicted in figure \ref{fig:Ramsey}b). 
To investigate this interference of the multimode output after scattering on the  two-level system, we use a similar approach as above. We place a delay cavity in each output path of the atomic decay channels and choose the delay cavity along the top path to pick up the radiation in a mode $w_1(t)$, while the delay cavity in the lower path picks up the mode $w_2(t)$. We can pick $w_1(t)$ and $w_2(t)$ as we desire, for instance, the modes that carry the highest average number of excitations. A method for finding these modes is described in \cite{long_kiilerich, short_kiilerich}. If we wish to investigate the multimode nature of the output field, we may pick up more modes using a number of delay cavities, but here we restrict ourselves to compute the output only in a single mode and treat all other modes as loss. Truncating the Hilbert space to only the relevant mode in question will yield the same results as if only the relevant mode is physically extracted for observation, e.g., by the quantum pulse gate \cite{sum-frequency-generation}.


For this example, we will analyse the output field from the non-linear scatterer in the same mode as the input modes. We therefore choose the delay cavities to pick up $w_1(t) = v_1(t)$ and $w_2(t)=v_2(t)$ respectively. We can then reemit them in the same shape, but with a new relative delay $\tau'$, if we desire to investigate their further propagation along another network, but we will not consider that here.

Introducing the two delay cavities after the atom will again increase the Hilbert space dimensionality,  but we will employ an interaction picture, which reduces the size of the Hilbert space and makes the problem tractable. The approach is very similar to the one in the previous section, and the derivation is deferred to the appendix.

We denote the operators of the delay cavities in the second interferometer as $\d_i$, where $\d_1$ is for the cavity in the long path and $\d_2$ for the short path. We also rename $\c_+ = \c_1$ and $\c_- = \c_2$ in the following to ease the notation. In the interaction picture with respect to the linear transfer between all these cavity modes, the operators are given by a time-dependent rotation of the initial operators at $t=0$
\begin{align}
    \c_i(t) &= \cos\theta_i(t) \c_i(0) - \sin\theta_i(t) \d_i(0) \\
    \d_i(t) &= \cos\theta_i(t) \d_i(0) + \sin\theta_i(t) \c_i(0),
\end{align}
where the rotation angle is given by
\begin{equation}
    \sin^2\theta_i(t) = \int_0^t dt' |v_i(t')|^2.
\end{equation}
The full Hamiltonian for this system in the interaction picture is $\H_I = \H_\textup{sys} + \H_\textup{int}$, where $\H_\textup{sys} = \Delta \hat{\sigma}_+ \hat{\sigma}_-$ and
\begin{align}
\begin{split}
    \H_\textup{int} &= i \sqrt{\frac{\gamma}{2}} \sum_{i=1}^2 v_i^*(t) \left(\c_i^\dagger \hat{\sigma}_- + \cot(2\theta_i(t)) \d_i^\dagger \hat{\sigma}_-\right) + \text{H.c.},
\end{split}
\end{align}
while dissipation operators in the interaction picture are similarly given by
\begin{align}
\begin{split}
    \L_{1,I} &= \sqrt{2} [- v_1(t) \csc(2\theta_1(t)) \d_1 - v_2(t) \csc(2\theta_2(t)) \d_2] \\&+ \sqrt{\gamma} \hat{\sigma}_-, \\
    \L_{2,I} &= \sqrt{2} [v_1(t) \csc(2\theta_1(t)) \d_1 - v_2(t) \csc(2\theta_2(t)) \d_2],
\end{split}
\end{align}
We solve the master equation with this Hamiltonian and these dissipation operators, to achieve the dynamics in the interaction picture. The atomic dynamics is the same in the Schrödinger picture and the interaction picture, but we notice that in the interaction picture, the $\c_i$-operators gradually rotate into the $\d_i$ operators, so the final population in the $\c_i$-oscillators in the interaction picture corresponds to a population in the $\d_i$-oscillators in the Schrödinger picture. Since the dynamics concerning the linear transfer of excitation between the cavities is handled by the interaction picture, the introduction of the two final delay cavities has simplified the dynamics, and this system is actually faster to compute numerically than the one without the final delay cavities \cite{interaction-picture}. This seems counter-intuitive but originates from the fact that the states in the cavities no longer explore the full Hilbert space, but only few states in the vicinity of the original photon number in the cavity at $t=0$, since the two-level system can only handle few excitations during the interaction. It is thus possible to treat large photon number states by truncating the Hilbert space to the few most relevant Fock states in the most populated modes in the interaction picture \cite{interaction-picture}.

\begin{figure}
    \centering
    \includegraphics[width=\linewidth]{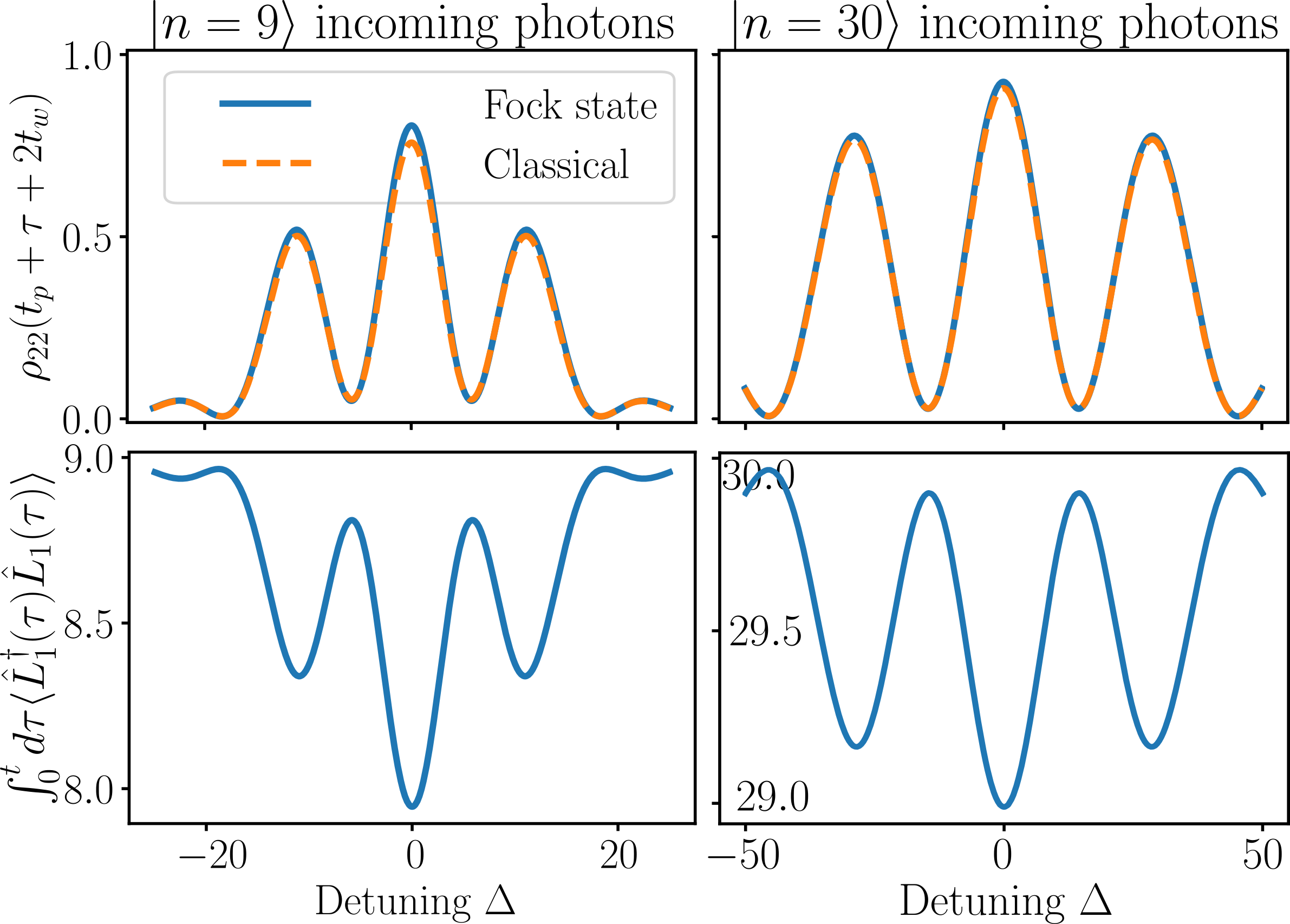}
    \caption{The blue, solid line in the two upper figures show the average atomic population as a function of detuning at two pulse widths after the peak of the second pulse at time $t = t_p + \tau + 2t_w$ for driving with a split-delayed Fock state as described in the main text. The population clearly show a Ramsey-like behaviour for both the $n=9$ and $n=30$ incoming photonic Fock state. The orange, dashed line in the same figures shows the atomic population driven by two equivalent classical pulses with average photon number equal to $n$. The two setups show the same qualitative behaviour. The lower panels show the integrated output intensity of the light in the initial mode $u(t)$ up to time $t = t_p + \tau + 2t_w$ in the output port where the photons would interfere constructively if there was no atom. The intensity clearly shows a similar behaviour as the atomic population, meaning that the population of the atom on average decays into orthogonal modes to the incoming mode. We make it clear that this does not mean the output Fock state is $\ket{n-1}$, only that the average number of photons approach $n-1$ in the zero-detuning case.}
    \label{fig:ramsey-photonic-intensity}
\end{figure}

The integrated intensity of the interference of the two delayed output pulses on a second interferometer for $\tau'=0$ is shown in figure \ref{fig:ramsey-photonic-intensity} as a function of detuning. The intensity of the output field displays a behaviour similar to the classical Ramsey-effect. At zero detuning, the atom interacts strongly with the incoming pulses, and the intensity in the pulse mode is reduced by one photon on average, while varying the detuning, a Ramsey-like interference pattern emerges. 

\section{Conclusion and outlook}
We have presented a theoretical method to handle pulse propagation delays in optical networks. Two methods have been presented: One which can delay a single pulse an arbitrary amount, and one that can delay any number of pulses relative to other quantum channels. We presented examples of the theory and its application to evaluate the outcome of exciting a two-level atom by two entangled quantum pulses of radiation. Further exploration of the method may deal with dispersive propagation, described by a modified shape of the delayed pulses, sensing with interferometers, manipulation with time bin qubits, and studying time-delayed interactions between quantum emitters.

V. Rueskov Christiansen acknowledges support from the Danish National Research Foundation through the Center of Excellence for Complex Quantum Systems (Grant agreement No. DNRF152).
K. Mølmer acknowledges support from the Danish National Research Foundation through the Hy-Q Center of Excellence for Hybrid Quantum Networks  (Grant agreement No. DNRF156).

\section{Appendix}
In this appendix, the interaction picture Hamiltonian and dissipation operators for the setup in section \ref{sec:non-linear-output} are derived, under the assumption that $w_i(t) = v_i(t)$. We take a starting point in the Hamiltonian and dissipation operators for an atom irradiated by a pulse which has traversed the Mach-Zehnder interferometer with different path lengths in section \ref{sec:Mach-Zehnder}, which is given by
\begin{align}
\begin{split}
    \H_1 = \Delta \hat{\sigma}_+ \hat{\sigma}_- + \frac{i}{2} \sqrt{\frac{\gamma}{2}}&\left(\gout{v_1}(t) \c_1^\dagger \hat{\sigma}_- \right. \\ &\left. + \gout{v_2}(t) \c_2^\dagger \hat{\sigma}_- - \text{H.c.}\right),
\end{split}
\end{align}
where we have introduced the operators $\c_1 = \frac{1}{\sqrt{2}}(\a_u + \b_2)$ and $\c_2 = \frac{1}{\sqrt{2}}(\a_u - \b_2)$, which also simplifies the dissipation operators
\begin{align}
\begin{split}
    \L_1 &= \gout{v_1}^*(t) \c_1 + \sqrt{\frac{\gamma}{2}} \hat{\sigma}_-, 
\end{split}\\
\begin{split}
    \L_2 &= \gout{v_2}^*(t) \c_2 + \sqrt{\frac{\gamma}{2}} \hat{\sigma}_-.
\end{split}
\end{align}
Appending delay cavities to the two output ports that pick up the radiation in the $v_1(t)$ and $v_2(t)$ modes 
results in the addition of the following terms to the Hamiltonian
\begin{align}
\begin{split}
    \H_2 = \H_1 &+ \frac{i}{2}\left(\sqrt{\frac{\gamma}{2}} \left(\gin{v_1}^*(t) \hat{\sigma}_+ \d_1 + \gin{v_2}^*(t) \hat{\sigma}_+ \d_2\right)\right. \\
    &\left. + \gout{v_1}(t)\gin{v_1}^*(t) \c_1^\dagger \d_1 \right. \\
    &\left.+ \gout{v_2}(t)\gin{v_2}^*(t) \c_2^\dagger \d_2 - \text{H.c.}\right).
\end{split}
\end{align}
The dissipation operators are
\begin{align}
\begin{split}
    \L_1 &= \frac{1}{\sqrt{2}} [\gin{v_1}^*(t) \d_1 + \gin{v_2}^*(t) \d_2] \\&+ \frac{1}{\sqrt{2}}(\gout{v_1}^*(t) \c_1 + \gout{v_2}^*(t) \c_2) + \sqrt{\gamma} \hat{\sigma}_-
\end{split} \\
\begin{split}
    \L_2 &= \frac{1}{\sqrt{2}} [\gin{v_2}^*(t) \d_2 - \gin{v_1}^*(t) \d_1] \\&+ \frac{1}{\sqrt{2}}(\gout{v_2}^*(t) \c_2 - \gout{v_1}^*(t) \c_1)
\end{split}
\end{align}
Now we go to the interaction picture with respect to the cavity-cavity part of the Hamiltonian
\begin{align}
\begin{split}
    \H_0 = \frac{i}{2}&\left(\gout{v_1}(t)\gin{v_1}^*(t) \c_1^\dagger \d_1 \right. \\
    &\left.+ \gout{v_2}(t)\gin{v_2}^*(t) \c_2^\dagger \d_2 - \text{H.c.}\right)
\end{split}
\end{align}
This leads to the following equations of motion for the interaction picture operators
\begin{align}
    \frac{d}{dt}\c_i = \frac{1}{2} \gout{v_i}(t)\gin{v_i}^*(t) \d_i, \\
    \frac{d}{dt}\d_i = -\frac{1}{2} \gin{v_i}(t) \gout{v_i}^*(t) \c_i,
\end{align}
for $i=1, 2$. The solution to these differential equations is the same as we saw in the main text
\begin{align}
    \c_i(t) &= \cos\theta_i(t) \c_i(0) - \sin\theta_i(t) \d_i(0) \\
    \d_i(t) &= \cos\theta_i(t) \d_i(0) + \sin\theta_i(t) \c_i(0),
\end{align}
where $\theta_i(t)$ is defined by
\begin{align}
    \sin^2\theta_i(t) = \int_0^t dt' |v_i(t')|^2.
\end{align}
This transformation gives that the Hamiltonian in the interaction picture is
\begin{align}
\begin{split}
    \H_I &= \Delta \hat{\sigma}_+ \hat{\sigma}_- \\
    &+ i \sqrt{\frac{\gamma}{2}} \sum_{i=1}^2 v_i^*(t) \left(\c_i^\dagger \hat{\sigma}_- + \cot(2\theta_i(t)) \d_i^\dagger \hat{\sigma}_- - \text{H.c.}\right),
\end{split}
\end{align}
and the dissipation operators become
\begin{align}
\begin{split}
    \L_{1,I} &= \sqrt{2} [- v_1(t) \csc(2\theta_1(t)) \d_1 - v_2(t) \csc(2\theta_2(t)) \d_2] \\ &+ \sqrt{\gamma} \hat{\sigma}_- 
\end{split}\\
\begin{split}
    \L_{2,I} &= \sqrt{2} [v_1(t) \csc(2\theta_1(t)) \d_1 - v_2(t) \csc(2\theta_2(t)) \d_2].
\end{split}
\end{align}

%

\end{document}